# Wear-resistant thin films of MRI-230D-Mg alloy using plasma-driven electrolytic oxidation

*G. Rapheal, S. Kumar, C. Blawert, Narendra B. Dahotre*

*Wear resistant coatings were produced on a permanent mould cast MRI 230D Mg alloy by (a) PEO in silicate based electrolyte, (b) PEO in phosphate based electrolyte, (c) hybrid coatings of silicate PEO followed by laser surface alloying (LSA) with Al and Al2O3, and (d) hybrid coatings of phosphate PEO followed by LSA with Al and Al2O3. Microstructural characterization of the coatings was carried out by scanning electron microscopy (SEM) and X(ray diffraction. The tribological behavior of the coatings was investigated under dry sliding condition using linearly reciprocating ball-on-flat wear test. Both the PEO coatings exhibited a friction coefficient of about 0.8 and hybrid coatings exhibited a value of about 0.5 against the AISI 52100 steel ball as the friction partner, which were slightly reduced with the increase in applied load. The PEO coatings sustained the test without failure at 2 N load but failed at 5 N load due to micro-fracture caused by high contact stresses. The hybrid coatings did not get completely worn off at 2 N load but were completely removed exposing the substrate at 5 N load. The PEO coatings exhibited better wear resistance than the hybrid coatings and silicate PEO coatings exhibited better wear resistance than the phosphate PEO coatings. Both the PEO coatings melted/decomposed on laser irradiation and all the hybrid coatings exhibited similar microstructure and wear behavior irrespective of the nature of the primary PEO coating or laser energies. SEM examination of worn surfaces indicated abrasive wear combined with adhesive wear for all the specimens. The surface of the ball exhibited a discontinuous transfer layer after the wear test.*

## 1. Introduction

Mg alloys are finding increasing applications due to their attractive combination of properties, such as, high specific strength, excellent die-castability, good machinability, vibration damping, electromagnetic interference shielding properties and recyclability [1]. Today, Mg alloys are widely used as die castings in automotive, aerospace and Information and Communication Technology sectors. During the last decade a great deal of research has been done on developing and optimizing new Mg alloys with improved properties, which are considered to be a direct replacement for Al alloys and ferrous materials. The weight reduction achieved as a result of direct replacement of parts with Mg alloys in automotive and aerospace sectors reflects in increased efficiency and lower green house gas emissions. However, a serious impediment in the successful application of Mg alloys is its low wear resistance [2].

A wide variety of surface coatings, viz., organic, conversion, CVD, PVD, plasma electrolytic oxidation (PEO), laser surface alloying (LSA), etc. have emerged as effective means to improve the wear resistance of Mg alloys [3–7]. Among these coating processes, LSA and PEO have gained attention during recent years due to the exhibited increased wear resistance [8–27]. PEO coatings are basically conversion coatings performed at high voltages in an aqueous electrolyte, which rely on repetitive local dielectric breakdown and formation of plasma modifying the coating with the incorporation of species from the electrolyte [28]. Coatings produced by PEO treatment consist of hard crystalline ceramic phases, which have good adherence to the substrate. However, they are porous and rough and exhibit high coefficient of friction under dry sliding conditions. The open and interconnected pore structure, in fact, makes them vulnerable to fracture failure under load and reduced corrosion resistance, especially in the long run. Among PEO coatings on light metals, two widely studied types are based on silicate and phosphate electrolytes. Lasers, due to their inherent property of yielding high power density, low total energy beams are particularly suitable for surface modifications processes, viz., melting, alloying, cladding, etc. [29].

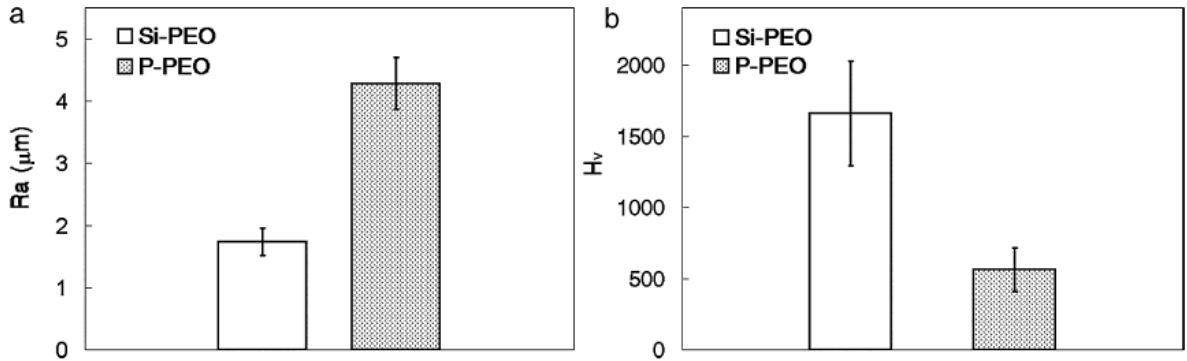

Fig. 1. (a) Average surface roughness and (b) Vickers hardness of PEO coatings.

In the present investigation, a permanent mould cast MRI 230D Mg alloy has been subjected to silicate and phosphate based PEO treatments as well as to hybrid treatments consisting of PEO and LSA with Al and Al2O3 in order to improve the wear resistance of the alloy. MRI 230D is a creep resistant Mg alloy developed for automotive power train applications by Magnesium Research Institute, a joint venture between Dead Sea Magnesium and Volkswagen AG [30–41].

## 2. Experimental procedure

Permanent mould cast MRI 230D alloy ingot of nominal composition (in wt.%) 6.45% Al, 2.25% Ca, 0.25% Sr, 0.84% Sn, 0.27% Mn, Zn, bal. Mg, was sliced into plates of size 50 mm × 50 mm × 5 mm and used as substrate for coatings. Specimens were ground successively with emery papers of 500, 800 and 1000 grit size, cleaned with ethyl alcohol followed by distilled water prior to PEO treatment. PEO coating was carried out in alkaline silicate and phosphate based electrolytes at a constant current density of 15 mA/cm2 for 30 min with the specimen serving as anode. Electrolytes were prepared using analytical grade reagents in double distilled water with a composition of 10 g/l Na2SiO3 + 1 g/l KOH for silicate and 10 g/l Na3PO4 + 1 g/l KOH for phosphate PEO coatings. The temperature of the electrolyte bath was maintained constant at 10 ± 2 ∘C by water circulation through a dedicated chiller unit. A custom built, variable duty cycle pulsed DC power supply capable of delivering 5A at 600V was used for the coatings. Duty cycle of the pulse employed for the coatings was 10% with ton = 2 ms and toff = 18 ms. Specimens were thoroughly rinsed in water immediately after the PEO treatment and dried in ambient conditions. Silicate and phosphate PEO coatings are henceforth designated as Si-PEO and P-PEO coatings. For the preparation of hybrid coatings by LSA of PEO treated specimens, pre-placed precursor method was adopted. First, precursor powder of Al (99.5% purity, 10 m particle size) was mixed with a previously prepared 95 vol% water and 5 vol% proprietary water-based organic solvent 'LISIW15853' (Warren Paint and Color Company, Nashville, TN, USA) and was spray deposited on the PEO treated plates to a thickness of 80 m and subjected to laser irradiation. Subsequently, the precursor powder consisting of Al2O3 (99.0% purity, 10 m particle size) was spray deposited to a thickness of 80 m and laser irradiated. Hybrid coatings are designated as (Si, P)(PEO + LSA (5, 5.5, 6, 6.5)J in the text, as the case may be. For laser irradiation, a 400W mean power LumonicsTM JK701 pulsed Nd:YAG laser equipped with fiber-optic beam delivery system was used. Laser beam was delivered with uniform energy distribution, both in spatial and temporal coherence, employing a 120 mm focal length convex lens, which provided a defocused spot diameter of 800 m on the specimen surface. Laser surface treatment was carried out at a scan speed of 42 mm/s with a pulse width of 0.5 ms, repetition rate of 20 Hz using argon as cover gas at a pressure of 5.5 bar. Laser pulse energies of 5, 5.5, 6 and 6.5 J, which correspond to beam energies 100, 110, 120 and 130W, were employed for the coating process. Several laser tracks were laid with minimal overlap to cover the entire surface area.

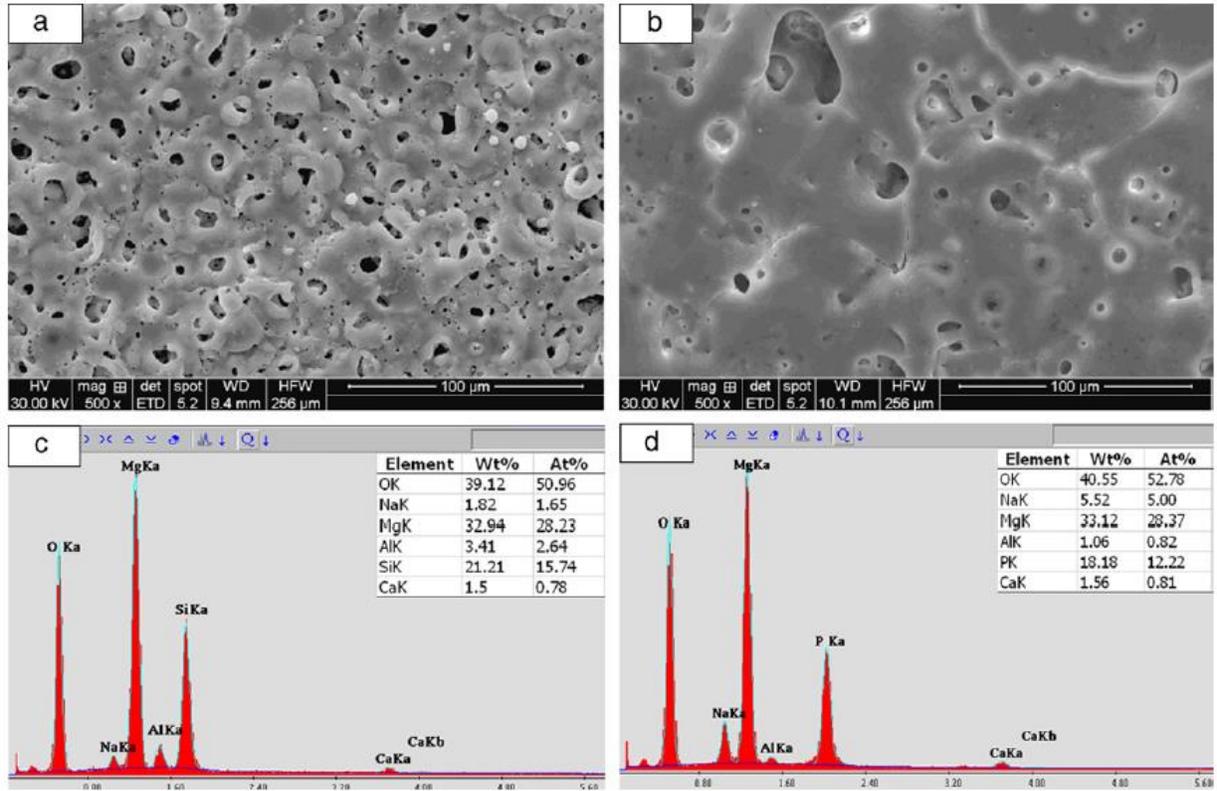

Fig. 2. SEM images of the top surface of PEO coatings (a) Si-PEO, (b) P-PEO, (c) EDS spectrum of Si-PEO and (d) EDS spectrum of P-PEO.

Surface roughness of PEO coatings was measured by HommelwerkeTM T1000 surface profile roughness gage and hardness was measured by CSMTM instrumented indentation test set-up using a load of 0.1 N. Friction and wear behavior of the coatings were analyzed by a Tribotec ball on disc oscillating tribometer (TribotechnicTM, 92110, Clichy, France) conforming to ASTM G-133 standard. AISI 52100 steel ball of diameter 6 mm was employed as the friction partner. Tests were conducted at a sliding speed of 5 mm/s for a total sliding distance of 100 m at normal loads of 2 N and 5 N with oscillating amplitude of 10 mm. Microstructural characterization of the surface and crosssection of the coatings were carried out on FEI QuantaTM 200 scanning electron microscope (SEM) equipped with Oxford InstrumentsTM energy dispersive X-ray spectroscopy (EDS) analysis. Standard metallographic techniques were employed for specimen preparation. Etching reagent employed was a solution comprising of 100 ml ethanol, 10 ml acetic acid, 6 ml picric acid and 20 ml distilled water. The phase composition of the coatings were determined by X-ray diffraction with Cu-K radiation (0.154060 nm) using PANlytical X'Pert ProTM diffractometer.

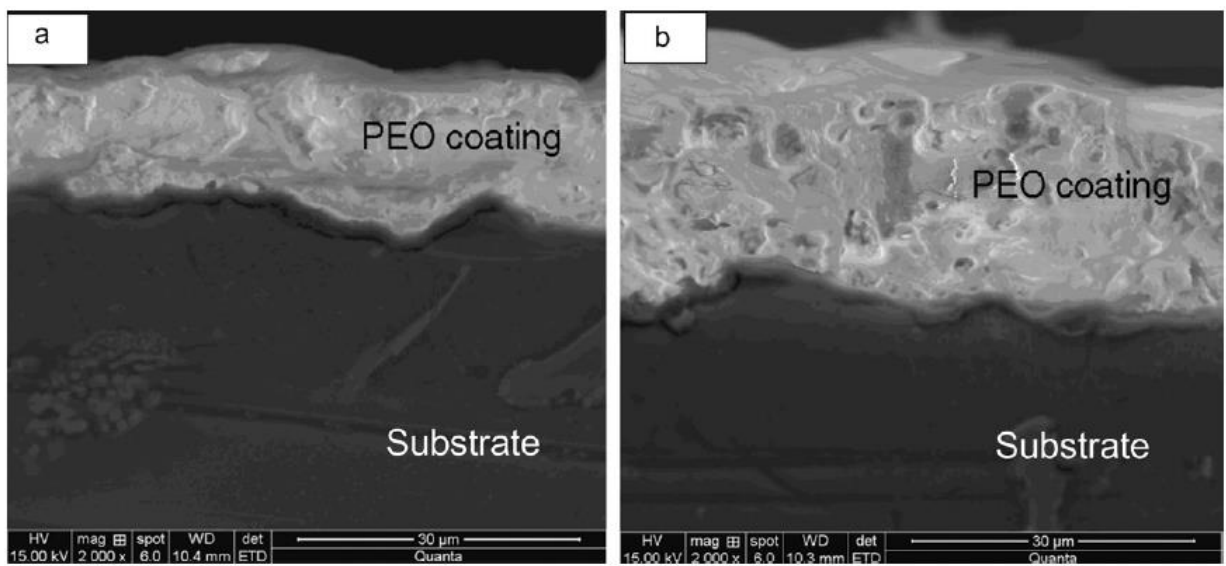

Fig. 3. SEM images of the cross section of PEO coatings (a) Si-PEO and (b) P-PEO.

## 3. Results and discussion

### 3.1. Microstructural characterization

Fig. 1(a) shows the average surface roughness of the PEO coatings. Si-PEO coating was relatively smooth with a Ra value of 1.7 ± 0.2 m, while P-PEO coating had a Ra value of 4.3 ± 0.4 m. The Si-PEO coating exhibited a much higher hardness of 1660 ± 367 Hv, almost three times the hardness of the P-PEO coating, i.e., 560 ± 153 Hv, as shown in Fig. 1(b). The ultra high hardness is due to the ceramic nature of these coatings. Fig. 2(a and b) shows the SEM image of the surface morphology of Si-PEO and P-PEO coatings. Both the coatings exhibited the characteristic porous feature of the PEO coatings. The Si-PEO specimen exhibited a large number of pores while the P-PEO specimen exhibited lesser number of pores, though they were larger in size. The larger size of pores on the surface of the coating may be the reason for the higher surface roughness of the P-PEO coatings observed in Fig. 1(a). Micro-cracks were also observed in the coatings, which can be attributed to thermal stresses generated by rapid solidification and cooling of ceramics produced by plasma discharges. EDS spectrum of Si-PEO and P-PEO coatings, as shown in Fig. 2(c and d), revealed the presence of Si and P respectively in the coatings. The incorporation of species from the electrolyte in the coating takes place by plasma-chemical reactions due to high energy release accompanying localized plasma discharges [45]. Fig. 3(a and b) shows the SEM images of cross-section of the Si-PEO and P-PEO coatings. The thickness of the Si-PEO coating was 11.5 ± 1.6 m while the thickness of the P-PEO coating was 17.3 ± 2.1 m. A few isolated pores were observed along the crosssection of the coatings as well, and they were found to be lesser in Si-PEO coating, which might also be responsible for their higher hardness than the P-PEO coating observed in Fig. 1(b). The interface ofthe coatings with substrate exhibited a wavy profile. The EDS line scan demonstrated that the elemental composition remained, in general, constant throughout the cross-section of the coating, as shown in Fig. 4. X-ray diffraction pattern of the PEO coatings is shown in Fig. 5. The Si-PEO coating essentially consisted of $Mg_2SiO_4$ with small proportions of MgO, $Al_2O_3$ and $Ca_2SiO_4$. The P-PEO coating mainly consisted of $Mg_3(PO_4)_2$ with small proportions of MgO. The difference in the composition of the two PEO coatings will contribute to the difference in hardness values observed in Fig. 1(b). Diffraction peaks corresponding to Mg were detected in both the PEO specimens and this may be due to the penetration of X-ray through the porous coating and subsequent diffraction from the substrate. SEM images of the cross-section of the hybrid coatings indicate good interfacial bonding with minimal heat affected zone (Fig. 6). A distinct layered structure of the hybrid coatings was not present in the coatings despite the dual coating process involved. This suggests that the primary PEO coatings along with the applied precursor and a part of the substrate melted/decomposed, mixed together and solidified on LSA treatment. A few solidification micro-cracks were visible in the hybrid coatings. A fine cellular microstructure was observed at higher magnification, as shown in the inset of Fig. 6(b). Alumina particles of different shapes and sizes were also observed in the coating. EDS line scan results on the cross-section of hybrid coatings are shown in Fig. 7. Mg concentration in hybrid coatings (Si-PEO + LSA 5.5J) gradually increased from ~75 at.% and reached the substrate value. Similarly, the concentration of Al gradually decreased from ~20 at.% and reached the substrate value. Si and P content were low in the hybrid coatings and appeared to be uniformly distributed throughout the crosssection of the coatings. As the scan traversed the grain boundary region, a dip in the concentration of Mg with simultaneous increase of Al and Ca was observed. This is due to the presence of the Laves phase C-36 [(Mg, Al)2Ca] at the grain boundary of the MRI 230D alloy [22–25]. Similar trends were observed in other hybrid coatings (both Si-PEO and P-PEO) processed with differentlaser energy; therefore, the microstructure and elemental composition profile did not vary with the change in laser beam energy in the range of 5–6.5 J.

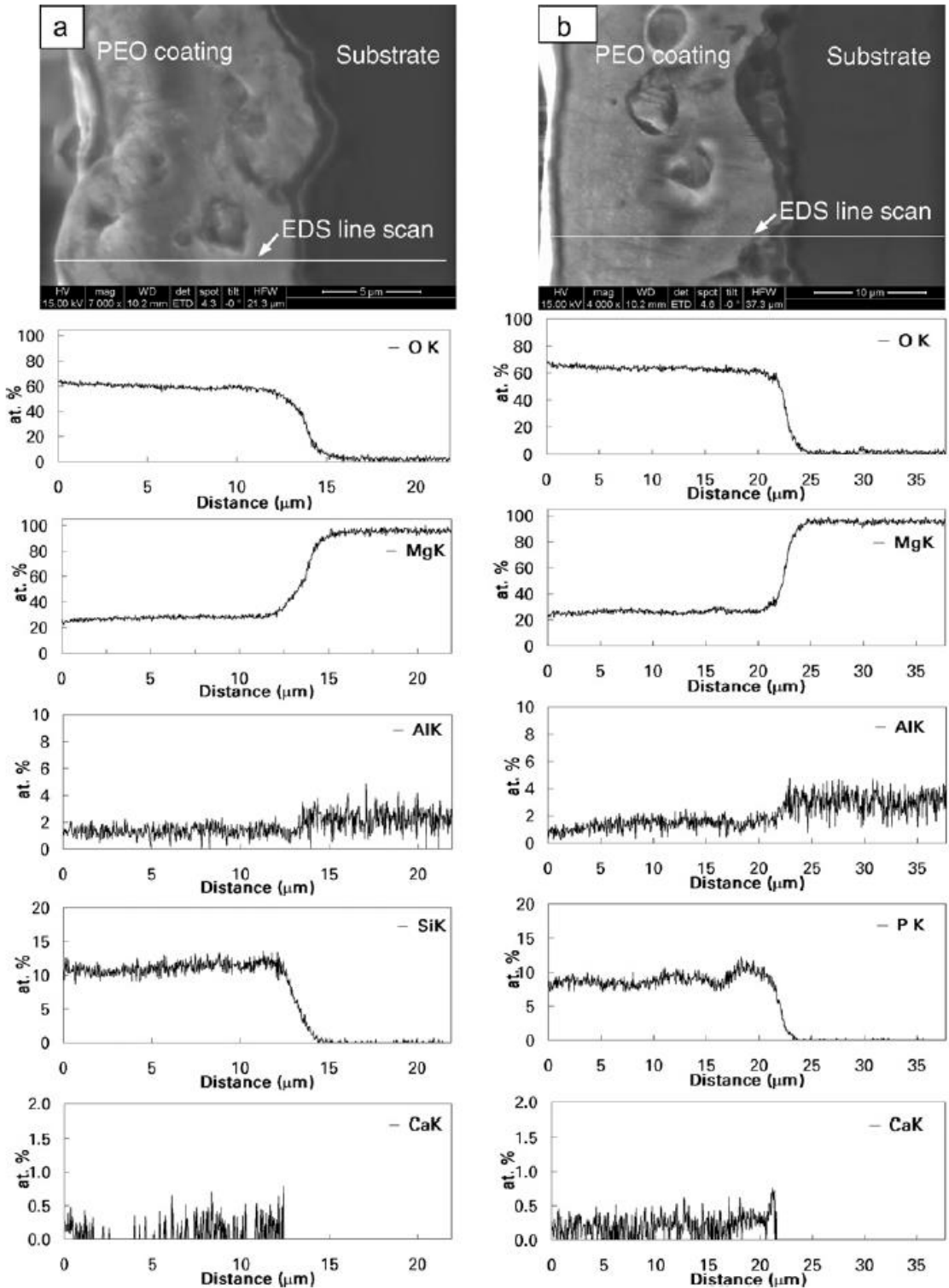

Fig. 4. EDS line scan across the cross-section of PEO coatings (a) Si-PEO and (b) P-PEO.

X-ray diffraction pattern of hybrid coatings is shown in Fig. 8. Both the silicate and phosphate based hybrid coatings consist of same phases, viz., -Mg (Mg17Al12), Al2O3 and MgO. Peaks corresponding to

Mg2SiO4 and Mg3(PO4)2 are absent in both the diffraction patterns, which confirms that the PEO coatings are completely melted/decomposed by laser beam. Similar diffraction patterns were obtainedfor specimensprocessedat other laser energies. Hardness profiles of the hybrid coatings exhibited similar trend (Fig. 9). An average value of 55 Hv was observed in the substrate region and hardness values were found to be gradually increasing towards the edge. Vickers hardness values of 273 Hv and 230 Hv were observed for (Si-PEO + LSA 5J) and (P-PEO + LSA 5J) hybrid coatings respectively at a distance of 25 m from the edge. The increased hardness appears to be due to the presence of hard - phase, solid solution strengthening, finer microstructure and the dispersion of alumina particles, as also reported by Hazra et al. [43].

### 3.2. Wear behavior

The plot of coefficient of friction against sliding distance for tests conducted at 2 N load for as-cast and PEO coated specimens is shown in Fig. 10(a). The as-cast material exhibited a steady state coefficient of friction of $0.37 \pm 0.03$ after an initial 16 m of sliding. The fluctuations in the coefficient offriction is due to periodic materialtransfer by adhesive wear from the specimen to the ball leading to the formation oftransfer layer and subsequentformation of wear debris by fracture of this transfer layer. The PEO coatings started with a coefficient of friction of ~0.4 but soon rose to much higher steady state values, i.e., $0.82 \pm 0.01$ after 13 m of sliding for the Si-PEO and $0.85 \pm 0.01$ after 44 m sliding distance for the P-PEO coating. The friction of ceramics against metallic materials, such as PEO coatings against steelinthepresent case,is accompaniedby the continual renovation of the friction contact area. Owing to this, the chemical interaction is accelerated in the contacting area, resulting in a rise in coefficient of friction [18,26–28]. These steady state values are maintained till the end of the test, which implies that these coatings survived the test without failure at 2 N load. The plot of coefficient of friction against sliding distance for tests conducted at 5 N load is shown in Fig. 10(b). The as-cast material exhibited a slightly lower steady state friction coefficient of $0.29 \pm 0.02$ at 5 N as compared to 2 N load. This is due to the relative ease in the establishment of adhesive transfer layer at higher loads, modifying the surfaces in contact and reducing the coefficient of friction.

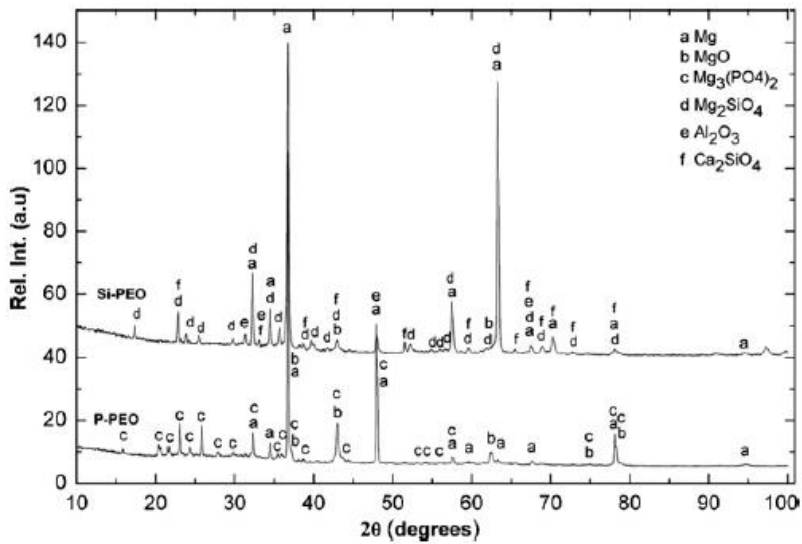

Fig. 5. X-ray diffraction pattern of Si-PEO and P-PEO coatings using Cu-K radiation ( 0.154060 nm).

The PEO coatings again exhibited a much higher initial coefficient of friction than the substrate. However, these values soon dropped to the level of as-cast material, which indicates the abrupt failure of these coatings at 5 N load. In case of P-PEO coatings the drop takes place immediately after the commencement of the test, whereas the SiPEO coatings survive up to about 20 m of sliding distance due to the higher strength of these coatings. SEM images of worn surface of as-cast material exhibited abrasive wear combined with severe plastic deformation characterized by the formation of ridged grooves at both the loads of 2 N and 5 N, as shown in Fig. 11. EDS analysis of the worn surface exhibited the presence of all the alloying elements in the MRI 230D alloy along with oxygen. SEM images of the worn surface of the PEO coatings at 2 N load are shown in Fig. 12. The Si-PEO coating exhibited a few parallel grooves and the coating survived the entire duration of the test. For P-PEO coating, local failure of coatings was observed at few places, but it did not lead to the lowering of the coefficient of friction.

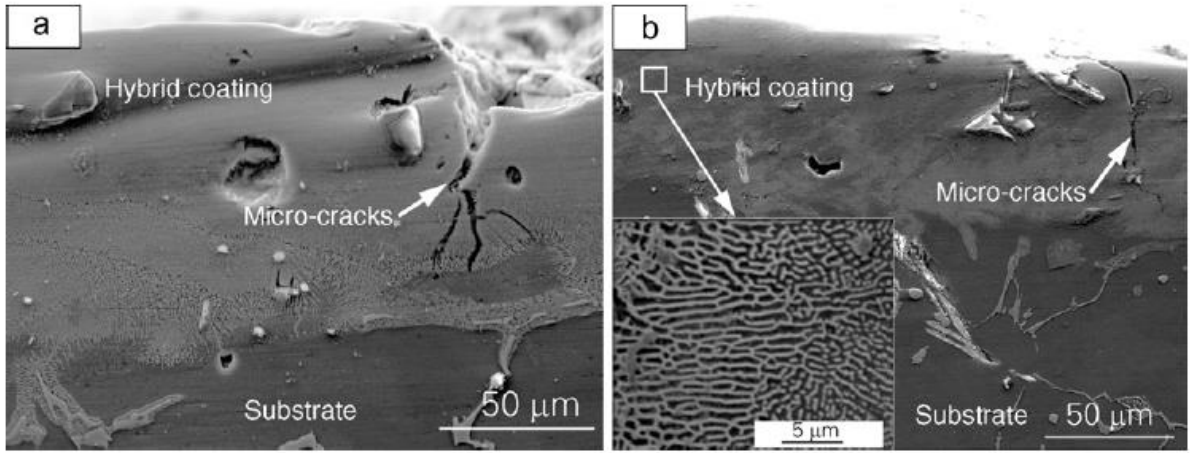

Fig. 6. SEM images of hybrid coatings (a) Si-PEO + LSA 5J and (b) P-PEO + LSA 5J.

The EDS analysis of the worn surfaces of both the Si-PEO and P-PEO coatings revealed the presence of Fe and O. Thus, the hard PEO coatings caused the abrasive wear of steel ball and the formation of oxide tribo-layer, which supports the argument put forth for the high steady state coefficient of friction in the previous paragraph. Srinivasan et al. have also reported abrasive wear of steel ball under dry sliding conditions for both Si-PEO and P-PEO coatings [47,49]. SEM images of the worn surface of the PEO coatings tested at 5 N load are shown in Fig. 13. A complete failure of coatings was observed in the micrographs. Micro-fracture of the coating due to high contact stresses is evident in the narrow band of partial coating failure between exposed substrate and the coating, as also observed by Srinivasan et al. [47,49]. The porous nature of the coatings, even though provides a great degree of compliance [50], is detrimentalto wear resistance. The EDS analysis ofthe worn surfaces of both the PEO coatings did not indicate any presence of Fe, and the Al content was close to that of the substrate, indicating the complete early failure of the coatings at 5 N load.

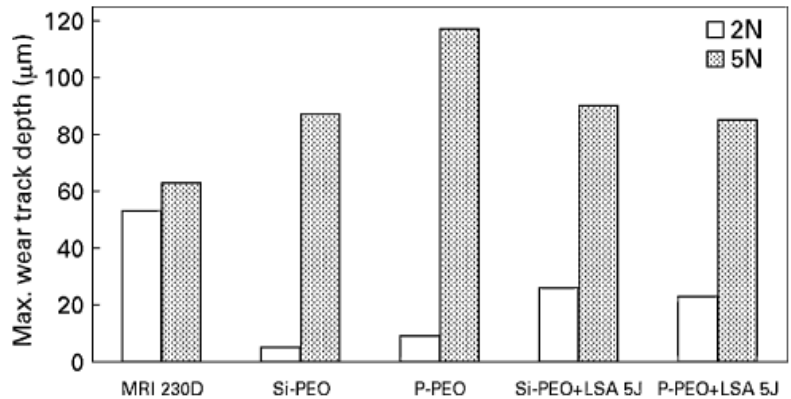

Fig. 19. Maximum wear track depth for as-cast, PEO and hybrid coatings.

## 4. Conclusions

Wear resistant surface coatings were produced on permanent mould cast MRI 230D Mg alloy by PEO treatment in alkaline silicate and phosphate based electrolytes, and by hybrid treatment of PEO and LSA with Al and Al2O3. The coatings were characterized for microstructure and its effect on dry sliding wear behavior using reciprocating steel ball-on-flat wear test was studied. The following conclusions can be drawn from the present investigation:

1. A characteristic porous coating was obtained by PEO treatment. A few micro-cracks due to thermal stresses were also observed on the surface of the coatings. The Si-PEO coating mainly consisted of Mg2SiO4 together with small amount of MgO, Al2O3 and Ca2SiO4, whereas the P-PEO coating mainly consisted of Mg3(PO4)2 together with small amount of MgO. Both the PEO coatings exhibited high hardness values, $1660 \pm 367$ for Si-PEO and $560 \pm 153$ Hv for P-PEO due to ceramicnature ofthe coatings.

2. A distinct layered coating was not observed in the hybrid coatings. The primary PEO coating was completely melted/decomposed along with applied precursor and substrate material, mixed together and re-solidified on LSA treatment with Al and Al2O3. A few micro-cracks due to thermal stresses were observed in the hybrid coatings. Both the hybrid coatings exhibited similar composition and hardness at all the laser energies employed. The hybrid coatings consisted of -Mg, (Mg17Al12), Al2O3, and MgO phases. The hybrid coatings exhibited hardness values of about 250 Hv and the increase in hardness is due to the presence of -phase, solid solution strengthening, microstructural refinement and dispersion of Al2O3 particles.

3. The coefficient of friction was highestfor the PEO coatings ($\sim$0.8) followed by hybrid coatings ($\sim$0.5) and substrate (0.37) at 2 N load, which slightly decreased at 5 N load. The Si-PEO coating sustained the wear test at 2 N load, while the P-PEO coating exhibited local failure in some regions. At 5 N load, both the PEO coatings failed due to micro-fracture caused by high contact stresses. However, the P-PEO coating failed immediately upon the commencement of test, whereas the Si-PEO coating failed after a sliding distance of 20 m. Hybrid coatings did not get completely worn off at 2 N load but were completely removed exposing the substrate at 5 N load.

4. Wear track depth measurements revealed improved wear resistance for both the PEO and hybrid coatings as compared to the as-cast material at 2 N load. PEO coatings exhibited better wear resistance than the hybrid coatings and Si-PEO exhibited better wear resistance than the P-PEO coatings. The better wear resistance is directly correlated to the higher hardness values of the respective specimens. At 5 N load, wear track depth of all the coatings were higher than the substrate. This is due to the abrasive three-body wear by hard ceramic wear debris entrapped in wear tracks in case of coated specimens. Laser energies employed in the present investigation did not have any effect on the wear behavior of the hybrid coatings.

5. The test specimens exhibited a combination of abrasive and adhesive wear. The friction partner (ball) showed the formation of a discontinuous transfer layer.

## References

[1]   A. K. Mondal, S. Kumar, C. Blawert, N. B Dahotre, *Effect of laser surface treatment on corrosion and wear resistance of ACM720 Mg alloy*, Surface and Coatings Technology, 2008, 202, 3187

[2]   Jeal, N. High-performance magnesium. *Adv. Mater. Process* 2005, *163*, 65–67.

[3]   Cam, G.; Kocak, M. Progress in joining of advanced materials. *Int. Mater. Rev* 1998, *43*, 1–44.

[4]   Aslani, M., Garner, C. M., Kumar, S., Nordlund, D., Pianetta, P., Nishi, Y., "Characterization of electronic structure of periodically strained graphene," Appl. Phys. Lett., 2015, 107 (18), 183507.

[5]   Schubert, E.; Klassen, M.; Zerner, I.; Walz, C.; Sepold, G. Light-weight structures produced by laser beam joining for future applications in automobile and aerospace industry. *J. Mater. Process. Technol* 2001, *115*, 2–8.

[6]   M. Hazara, A. K. Mondal, S. Kumar, C. Blawert, N. B Dahotre, *Laser surface cladding of MRI 153M magnesium alloy with (Al+ Al 2 O 3)*, Surface and Coatings Technology, 2009, 203, 2292

[7]   Lanza, M.; Lauro, A.; Scanavino, S. Fabrication and weldability in structures. *AL Alumin. Alloys* 2001, *13*, 80–86.

[8]   Haferkamp, H.; Niemeyer, M.; Dilthey, U.; Trager, G. Laser and electron beam welding of magnesium materials. *Weld. Cutt* 2000, *52*, 178–180.

[9]   Kumar, S, et. al., "Sequential electronic and structural transitions in VO2 observed using x-ray absorption spectromicroscopy," 2014, Advanced Materials, 26, 7505.

[10]  Sanders, P.G.; Keske, J.S.; Leong, K.H.; Kornecki, G. High power Nd:YAG and CO₂ laser welding of magnesium. *J. Laser. Appl* 1999, *11*, 96–103.

[11]  Jayamurugan, G, Vasu, K. S., Rajesh, Y. B. R. D., Kumar, S., Vasumathi, V., Maiti, P. K., Sood, A. K., Jayaraman, N., "Interaction of single-walled carbon nanotubes with poly (propyl ether imine) dendrimers,", J. Chem. Phys. 2011, 134, 104507.

[12]  Leong, K.H.; Kornecki, G.; Sanders, P.G.; Keske, J.S. Laser Beam Welding of AZ31B-H24 Magnesium Alloy, Proceedings of the ICALEO 98: Laser Materials Processing Conference, Orlando, FL, USA, 16–19 November 1998; pp. 28–36.

[13]  Gerke, A, Kumar S., Provine, J., Saraswat K., "Characterization of metal-nitride films deposited by the Savannah ALD system," Stanford University.

[14]  Mordike, B.L.; Ebert, T. Magnesium: Properties-applications-potential. *Mater. Sci. Eng. A* 2001, *302*, 37–45.

[15]  Kumar, S, et. al., "The phase transition in VO2 probed using x-ray, visible and infrared radiations," Applied Physics Letters, 2016, 108, 073102

[16]  Marya, M.; Olson, D.L.; Edwards, G.R. Welding of Magnesium Alloys for Transportation Applications, Proceedings of the Materials Solution '00 on Joining of Advanced and Specialty Materials, St. Louis, MO, USA, 9–11 October 2000; pp. 122–128.

[17]  Suhas, K, "Materials and processes in 3D structuration of low temperature cofired ceramics for meso-scale devices," Industrial Ceramics, 2010, 30(1).

[18]  Pastor, M.; Zhao, H.; DebRoy, T. Continuous wave-Nd:yttrium–aluminium–garnet laser welding of AM60B magnesium alloys. *J. Laser. Appl* 2000, *12*, 91–100.

[19]  Kumar, S., "Energy from radioactivity," 2011, Stanford University

[20]  Aghion, E.; Bronfin, B. Magnesium alloys development towards the 21st century. *Mater. Sci. Forum* 2000, *350–351*, 19–30.

[21]  Suhas, K, Sripadaraja, K, "Mechanical modeling issues in optimization of dynamic behavior of RF MEMS switches," 2008, Int. J. Comp. Inform. Syst. Sci. Eng., 221-225.

[22]  Marya, M.; Edwards, G.; Marya, S.; Olson, D.L. Fundamentals in the Fusion Welding of Magnesium and Its Alloys, Proceedings of the Seventh JWS International Symposium, Kobe, Japan, 20–22 November 2001; pp. 597–602.

[23]    Kumar, S., "On the Thermoelectrically Pumped LED," 2012, Stanford University

[24]    Suhas, K., "Modelling Studies and Fabrication Feasibility with LTCC for Non-Conventional RF MEMS Switches," 2007, IISC.

[25]    Cao, X.; Jahazi, M.; Immarigeon, J.P.; Wallace, W. Review of laser welding techniques for magnesium alloys. *J. Mater. Process. Technol* 2006, *171*, 188–204.

[26]    Kumar, S., "Fundamental limits to Moore's law," 2012, Stanford University

[27]    Liu, P.; Li, Y.J.; Geng, H.R.; Wang, J. Microstructure characteristics in TIG welded joint of Mg Al dissimilar materials. *Mater. Lett* 2007, *61*, 1288–1291.

[28]    Baker, H.; Okamoto, H. *Alloy Phase Diagrams*, 9th ed; ASM International: Geauga County, OH, USA, 1995; Volume 3.

[29]    Kumar, S, et. al., "Characterization of electronic structure of periodically strained graphene," Applied Physics Letters, 2015, 107, 183507.

[30]    Wang, J.; Li, Y.J.; Liu, P.; Geng, H.R. Microstructure and XRD analysis in the interface zone of Mg-Al diffusion bonding. *J. Mater. Process. Technol* 2008, *205*, 146–150.

[31]    Kumar, S., Esfandyarpour, R., Davis, R., Nishi, Y., "Charge sensing by altering the phase transition in VO2," APS March Meeting Abstracts 2014, 1, 1135.

[32]    Hajjari, E.; Divandari, M.; Razavi, S.H.; Emami, S.M.; Homma, T.; Kamado, S. Dissimilar joining of Al Mg light metals by compound casting process. *J. Mater. Sci* 2011, *46*, 6491–6499.

[33]    Kumar, S., Esfandyarpour, R., Davis, R., Nishi, Y., "Surface charge sensing by altering the phase transition in VO2," Journal of Applied Physics 2014, 116, 074511.

[34]    Miyashita, Y.; Borrisutthekul, R.; Chen, J.; Mutoh, Y. Application of twin laser beam on AZ31-A5052 dissimilar metals welding. *Key Eng. Mater* 2007, *353–358*, 1956–1959.

[35]    Kumar, S, "Learning from Solar Cells - Improving Nuclear Batteries,", 2012, Stanford University,

[36]    Firouzdor, V.; Kou, S. Al-to-mg friction stir welding effect of material position, travel speed, and rotation speed. *Metall. Mater. Trans. A* 2010, *41*, 2914–2935.

[37]    Kumar, S, "Superconductors for electrical power," 2011, Stanford University

[38]    Yutaka, S.S.; Seung, H.C.P.; Masato, M.; Hiroyuki, K. Constitutional liquation during dissimilar friction stir welding of Al and Mg alloys. *Scr. Mater* 2004, *50*, 1233–1236.

[39]    Kumar, S., "Types of atomic/nuclear batteries," 2012, Stanford University

[40]    Hirano, S.; Okamoto, K.; Doi, M.; Okamura, H.; Inagaki, M.; Aono, Y. Microstructure of dissimilar joint interface of magnesium alloy and aluminum alloy by friction stir welding. *Quart. J. Jpn. Weld. Soc* 2003, *21*, 539.

[41]    Kumar, S, "Mechanisms of resistance switching in various transition metal oxides," Thesis, Stanford University (2014)

[42]    Park, S.H.C.; Michiuchi, M.; Sato, Y.S.; Kokawa, H. Dissimilar Friction-Stir Welding of Al Alloy 1050 and Mg Alloy AZ31, Proceedings of the International Welding/Joining Onference-Korea, Gyeongju, Korea, 28–30 October 2002; pp. 534–538.

[43]    Kumar, S, et. al., "Smart packaging of electronics and integrated MEMS devices using LTCC," 2016, arXiv:1605.01789

[44]    Yan, Y.; Zhang, D.T.; Qiu, C.; Zhang, W. Dissimilar friction stir welding between 5052 aluminum alloy and AZ31 magnesium alloy. *Trans. Nonferrous Met. Soc. China* 2010, *20*, S619–S623.

[45]    Kumar, S, et. al., 2013, "Local temperature redistribution and structural transition during Joule-heating-driven conductance switching in VO2," Advanced Materials, 2012,25, 6128.

[46]    Somasekharan, A.C.; Murr, L.E. Microstructures in friction-stir welded dissimilar magnesium alloys and magnesium alloys to 6061-T6 aluminum alloy. *Mater. Charact* 2004, *52*, 49–64.

[47]    Kumar, S., "Atomic Batteries: Energy from radioactivity," 2015, arXiv:1511.07427, http://arxiv.org/abs/1511.07427

[48]    Straumal, B.B.; Baretzky, B.; Kogtenkova, O.A.; Straumal, A.B.; Sidorenko, A.S. Wetting of grain boundaries in Al by the solid Al$_3$Mg$_2$ phase. *J. Mater. Sci* 2010, *45*, 2057–2061.

[49]    Kumar, S., 2015, "Fundamental limits to Morre's law", arXiv:1511.05956, http://arxiv.org/abs/1511.05956

[50]    Protasova, S.G.; Kogtenkova, O.A.; Straumal, B.B.; Zieba, P.; Baretzky, B. Inversed solid-phase grain boundary wetting in the Al–Zn system. *J. Mater. Sci* 2011, *46*, 4349–4353.